\documentclass[12pt,preprint]{aastex}

\def\pc{{\rm pc}}

\def\au{{\rm AU}}
\def\min{{\rm min}}

\def\habit{{\rm habit}}

\begin{document}

\title{Early-type Stars: Most Favorable Targets for Astrometrically 
Detectable Planets in the Habitable Zone}

\author 
{Andrew Gould}
\affil{Ohio State University, Department of Astronomy, Columbus, OH
43210}
\email{gould@astronomy.ohio-state.edu} 
\and
\author 
{Debra A.\ Fischer}
\affil{University of California, Department of Astronomy, Berkeley, CA
94720}
\email{fischer@astron.berkeley.edu}

\singlespace

\begin{abstract}

Early-type stars appear to be a difficult place to look for
planets astrometrically.  First, they are relatively heavy, and for
fixed planetary mass the astrometric signal falls inversely as 
the stellar mass.  Second, they are relatively rare (and so tend to
be more distant), and for fixed orbital separation the astrometric
signal falls inversely as the distance.  Nevertheless, because
early-type stars are relatively more luminous, their habitable zones
are at larger semi-major axis.  Since astrometric signal scales
directly as orbital size, this gives early-type stars a strong advantage,
which more than compensates for the other two factors.  Using the
Hipparcos catalog, we show that early-type stars constitute the majority 
of viable targets for astrometric searches for planets in the habitable
zone.  We contrast this characteristic to transit searches, which are 
primarily sensitive to habitable planets around late-type stars.

\end{abstract}
\clearpage
 
\section{Introduction
\label{sec:intro}}

To date, extrasolar planets have been discovered by three methods:
pulsar timing \citep{wolz}, radial velocities (RV, \citealt{mayor}), and
transits \citep{ogle,confirm}.  While RV has been by far the most successful
of these
\citep{butler}, it appears to be ultimately limited to $1\,\rm m\,s^{-1}$ 
precision 
by instabilities in the atmospheres of stars.  For planets in $a\sim 1\,\au$
orbits around solar-type stars, this corresponds to a planetary mass
$m_p\sim 10\,M_\oplus$.  Hence, to find terrestrial planets will probably
require other techniques.

Of particular interest are terrestrial planets in the so-called ``habitable
zone''.  While the exact specifications of this concept are the subject
of continuing study and debate, for purposes of this paper, we will
assume the habitable zone to be centered at
\begin{equation}
a_\habit = 1\,\au\,\sqrt{L\over L_\odot},
\label{eqn:ahabit}
\end{equation}
where $L$ is the bolometric luminosity of the star.

A quick survey of various techniques reveals four with reasonable potential
for detecting terrestrial planets: pulsar timing, microlensing 
\citep{mao},
transits, and astrometry \citep{shao}.  However, only two of these, transits
and astrometry, show reasonable prospects of detecting habitable planets.
Pulsar timing has already found three terrestrial planets in one system, but
the intense radiation field from the pulsar makes these almost certainly
uninhabitable.  Microlensing is probably the most efficient method for
detecting Earth mass (and even sub-Earth mass) planets \citep{bennett}.
However, it is sensitive primarily to planet-star separations
of an Einstein ring \citep{gl}, which is roughly at 
$a\sim 4\,\au (M/M_\odot)^{1/2}$, where $M$ is the mass of the star.  This
is well outside the habitable zone.

Detection of terrestrial planets, including in the habitable zone, is
the primary goal of three proposed space missions:
{\it Kepler}\footnote{http://www.kepler.arc.nasa.gov/} and
{\it Eddington}\footnote{http://sci.esa.int/home/eddington/index.cfm}
plan to search for these by planetary transits, while the
{\it Space Interferometry 
Mission (SIM)}\footnote{http://planetquest.jpl.nasa.gov/SIM/sim\_index.html}
plans to search for them from the astrometric wobble they induce on
their parent star.  

\citet{gpd} showed that the transit technique is actually most sensitive
to habitable planets orbiting low-mass stars because their low luminosity 
moves the habitable zone inward (see eq.\ [\ref{eqn:ahabit}]) where
transits are most efficiently detected.  This factor, together with the
greater abundance and smaller radii of low-mass stars, more than compensate 
for the worse signal-to-noise ratio (S/N) due to their lower brightness.
Here we ask whether the particularities of the astrometry  also
influence which stellar population  this technique is most sensitive to 
when searching for habitable planets.

\section{Analytic Investigation
\label{sec:analytic}}

Consider a homogeneous population of stars of mass $M$,
bolometric luminosity $L$, and space density $n$.  Assume that
astrometric wobbles of amplitude $\alpha_\min$ can be reliably detected.
A planet of mass $m_p$ in the habitable zone (as defined by
eq.\ [\ref{eqn:ahabit}]), can then be detected out to a distance
$d = (m_p/M)(L/L_\odot)^{1/2}(\au/\alpha_\min)$.  Hence,  
the total number of systems that can be probed is,
$$
N = {4\pi\over 3}n
\biggl({m_p \over M}\biggr)^3
\biggl({L\over L_\odot}\biggr)^{3/2}
\biggl({\alpha_\min \over\au}\biggr)^{3}
$$
\begin{equation}
= 7.6\,
{n\over n_0}\,
\biggl({m_p \over 3\,M_\oplus}\biggr)^{3}
\biggl({M \over M_\odot}\biggr)^{-3}
\biggl({L\over L_\odot}\biggr)^{3/2}
\biggl({\alpha_\min \over 1\,\mu\rm as}\biggr)^{-3},
\label{eqn:Neval}
\end{equation}
where we have normalized the density to $n_0=2.5\times 10^{-3}\,\pc^{-3}$,
the space density per $M_V$-magnitude of solar-type stars.  The key point
to note is that, in the neighborhood of $M\sim 1\,M_\odot$, the bolometric
luminosity scales as $L\propto M^\beta$, 
where $\beta\sim 4.5$.  Hence the mass 
and luminosity terms can be combined to yield
\begin{equation}
N\propto n M^{1.5\beta -3} \sim n M^{3.75}.
\label{eqn:Neval2}
\end{equation}
Although the number density of stars falls rapidly as a function of mass,
both because fewer are formed and those that do form live shorter lives,
this is somewhat compensated by the fact that these younger stars have
a lower scaleheight $h(M)$ and so are more concentrated near the plane.
The shorter lifetime scales as $t\propto M/L$, but this is relevant only
for stars $M\ga M_\odot$, whose lifetimes are shorter than the age of the
Galactic disk.  
The falling mass function
scales as $d n/d\ln M \propto M^{-x}$, where $x=1.35$ is the Salpeter value.
These factors can be combined to obtain,
\begin{equation}
{d N\over d\ln M} \propto  {M^{\beta/2 - x - 2}\over h(M)}
\sim {M^{-1.1}\over h(M)} \qquad M\ga M_\odot
\label{eqn:Nheavy}
\end{equation}
and,
\begin{equation}
{d N\over d\ln M} \propto  {M^{1.5\beta - x - 3}}
\sim {M^{2.4}} \qquad M\la M_\odot
\label{eqn:Nlight}
\end{equation}
Since the scaleheight factor in equation (\ref{eqn:Nheavy}) tends to counter
the mass factor, one expects that astrometric sensitivity will peak at
$M\sim M_\odot$ and then be roughly flat toward higher masses.

\section{Numerical Evaluation
\label{sec:numerical}}

\begin{figure}
\plotone{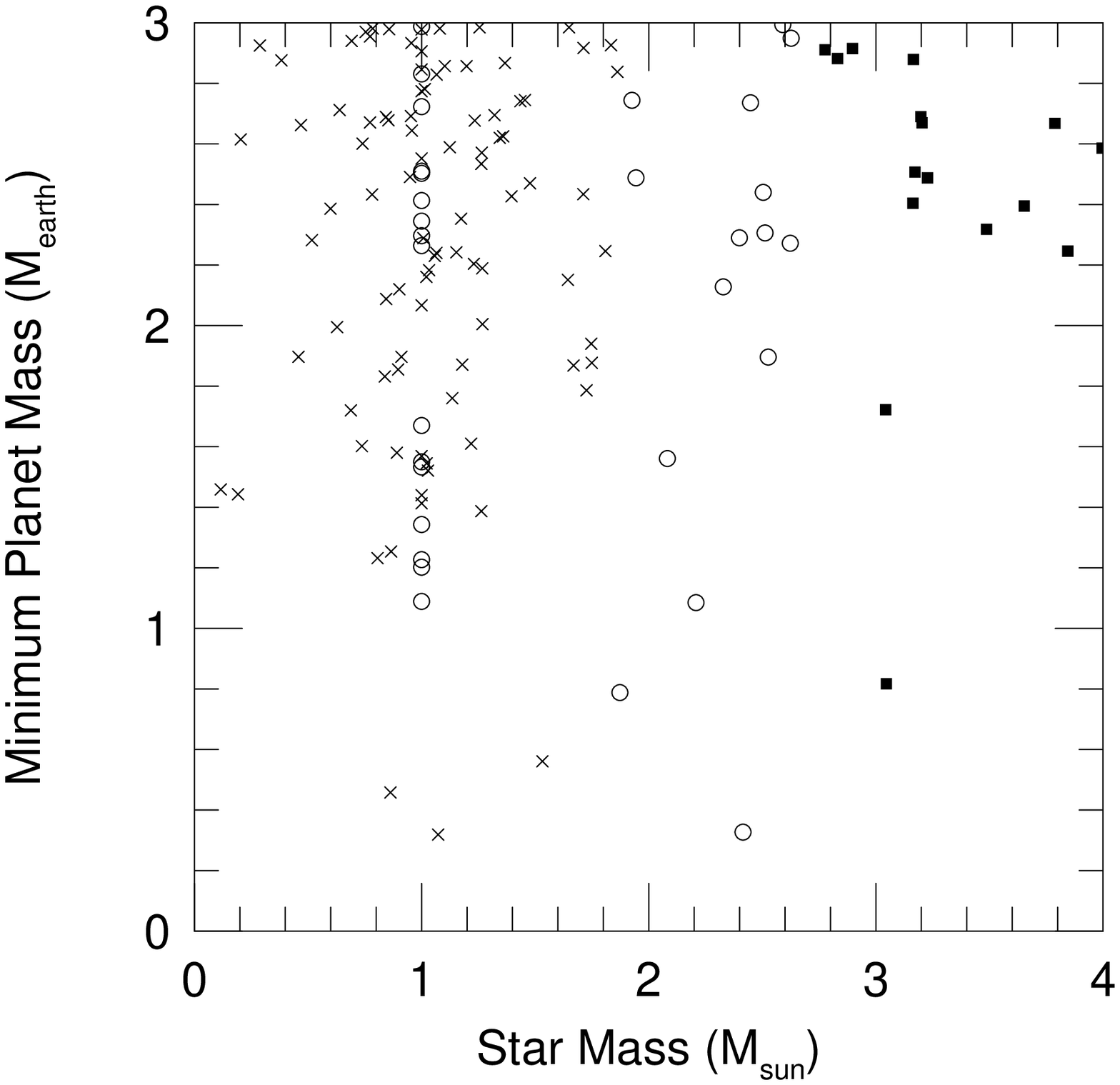}
\caption{\label{fig:minmass}
Minimum mass of detectable planets in the habitable zone as a function of 
stellar mass for stars in the solar neighborhood taken from the Hipparcos 
catalog \citep{hip}.  Planets with periods $P<5\,$yr ({\it crosses}),
$5\,{\rm yr}<P<10\,{\rm yr}$ ({\it open circles}), and 
$P>10\,$yr ({\it solid squares}), are shown separately.  Red giants are
all assigned a mass $M=1\,M_\odot$, but the $P>10\,{\rm yr}$ giants are not 
shown to avoid clutter.  For $P<5\,$yr, stars with $M>M_\odot$ outnumber
those with $M<M_\odot$ by 47 to 36, while for $P<10\,$yr the ratio is
62 to 36.
}\end{figure}

To obtain more definite estimates, we determine the minimum mass that
can be detected for each star in the Hipparcos catalog \citep{hip},
assuming that each has a planet with semi-major axis as given by 
equation (\ref{eqn:ahabit}).  We determine $M_V$ using the
Johnson $V$ mag and parallax as given by the Hipparcos catalog, and we find 
the bolometric corrections by making use of the
catalog's $V-I$ colors.  We estimate the mass using
$M_V$ and the mass-luminosity relation of \citet{allen}.  We adopt 
$\alpha_\min=1\,\mu\rm as$, which is the current best estimate
for $5\,\sigma$ detections for a 5-year {\it SIM} mission
(M.\ Shao 2002, private communication).  It corresponds to 50 measurements,
each with $1\,\mu\rm as$ precision for face-on orbits or 100 measurements for
edge-on orbits.  We also allow that this
performance will be a weak function of apparent magnitude,
$\alpha_\min = (1 + 10^{0.4(11-V)})^{1/2}\,\mu\rm as$, which has the
correct form in both the systematics-limited and photon-limited regimes.
However, including the flux-dependent term has hardly any effect on the 
results.  In addition, we arbitrarily
assign all red giants (defined as $M_V<4$ and $V-I>0.7$) a mass 
$M= 1.00\, M_\odot$. Red-giant masses are difficult to estimate, but
this estimate is not likely to be far off in most cases.  In addition,
fixing the red-giant masses at a unique value makes them easy to identify 
in Figure \ref{fig:minmass}, which displays our results.

	Planets with periods $P<5\,$yr
are shown by crosses, those with $P>10\,$yr are shown by solid squares,
and those in between by open circles.  However, the red giants 
$(M\equiv 1\,M_\odot)$ with $P>10\,$yr (of which there are 401) are
not shown to avoid clutter.  These period distinctions are important
because at the $5\,\sigma$ S/N limit used here, planets cannot be reliably
detected in less than one orbit.

Restricting consideration to planets with periods $P<5\,$yr, there are 
36 stars that can be probed with $M<M_\odot$ and 47 with $M>M_\odot$.
In addition, there are 9 red giants.  Some consideration is being given
to extending the {\it SIM} mission from 5 to 10 years\footnote{
The main reasons for an extended mission would be to push the sensitivity
to lower masses by increasing the S/N as well as to find longer-period planets.
For clarity, however, in Figure \ref{fig:minmass} we use the same 
$1\,\mu\rm as$ precision based on a 5-year mission for all stars 
regardless of period.
}.  
For this case,
the respective numbers are 36, 62, and 25.  In either case, the early-type
stars dominate over the late-type.  

\section{Discussion
\label{sec:discuss}}

\begin{figure}
\plotone{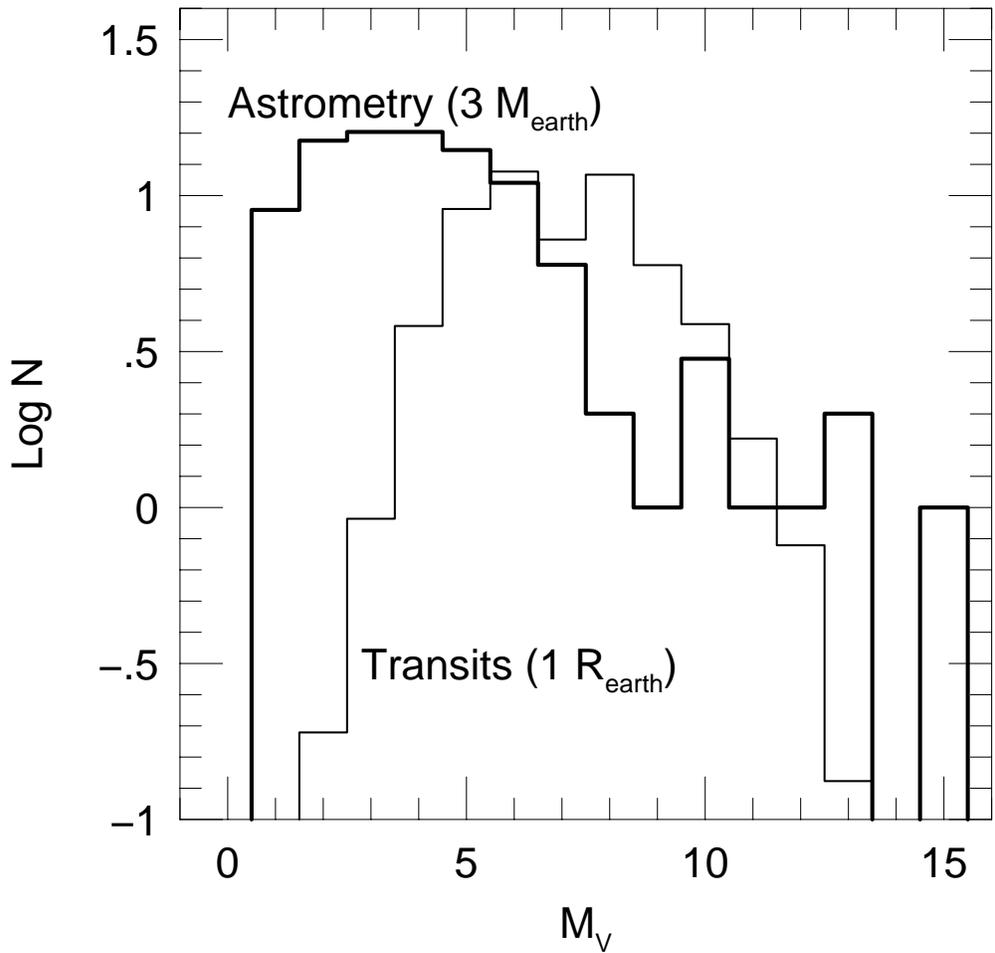}
\caption{\label{fig:mvhist}
Histograms of sensitivities of astrometric and transit surveys to planets
in the habitable zone.  The astrometric curve shows the same stars as in
Fig.~\ref{fig:minmass}, excluding all red giants and all stars with $P>10\,$yr.
The transit curve is taken from \citet{gpd}, who adopted the characteristics 
of the {\it Kepler} mission, 
but (contrary to the {\it Kepler} website) assumed 
that faint late-type dwarfs would be included in the survey.  Since astrometric
surveys are sensitive to mass, while transit surveys are sensitive to radius,
the two curves are not strictly comparable, but the relative trends with 
stellar type are robust.
}\end{figure}

The sensitivity of astrometric surveys to habitable planets around early-type
stars makes them complementary to transit surveys.  If these are properly
executed \citep{gpd}, they will be sensitive primarily to habitable
planets around late-type stars.  See Figure \ref{fig:mvhist}.

While early-type stars may have planets in the habitable zone, that
does not necessarily mean that these can be inhabited by natural processes.
On Earth, almost a Gyr was required before life took hold, and several
more were required before it succeeded in radically transforming the
atmosphere.  Very early stars do not live long enough to permit this
leisurely development.  However, there are a large number of targets
with masses $M\la 1.5\,M_\odot$, and these can live for $\ga 2.5\,$Gyr.
Moreover, it may be that life elsewhere in the universe develops faster
than on Earth.  Targeting these early-type stars for habitable-planet
searches is the best way to find out.

The time that red giants spend at approximately their current luminosity
(and so that a planet at a given semi-major axis would remain in the
habitable zone) is even shorter than the lifetime of many early-type stars.  
This would appear to give even less time for life to develop.  However, if
these stars had harbored planets in the habitable zone during their
main-sequence phase, and if these planets gave rise to intelligent,
technologically advanced life, these beings could have moved their home 
to progressively greater semi-major axes by orchestrating suitable
interactions with an asteroid or comet, in order to maintain their
planet's habitability as their star aged \citep{kla}.

\acknowledgments 
Work by AG was supported by JPL contract 1226901.

\clearpage

\end{document}